\documentclass[aps,prl,floatfix,showpacs,twocolumn, final
]{revtex4}

\usepackage{amsmath}
\usepackage{graphicx}

\usepackage{amsfonts}

\usepackage{epstopdf}

\newcommand{\bra}[1]{\mbox{$\langle #1|$}}
\newcommand{\ket}[1]{\mbox{$|#1\rangle$}}

\newcommand{\ketbra}[2]{\mbox{$|#1\rangle\langle #2|$}}

\begin{document}

\title{A photon loss tolerant Zeno CSIGN gate}

\author{Casey R. Myers$^1$ and Alexei Gilchrist$^{2*}$}
\affiliation{$^1$Department of Physics and Astronomy, Institute for Quantum Computing, University of Waterloo, ON, N2L 3G1, Canada. \\
$^2$Centre for Quantum
Computer Technology, Department of Physics, University of Queensland, QLD 4072, Brisbane,
Australia.
%, alexei@physics.uq.edu.au.
}

\date{\today}

\begin{abstract}
We model an optical implementation of a \textsc{csign} gate that makes use of
the Quantum Zeno effect \cite{04franson062302, 06franson053817} in the presence of
photon loss. The raw operation of the gate is severely affected by this type
of loss. However, we show that by using the same photon loss codes that have
been proposed for linear optical quantum computation (LOQC), the performance is
greatly enhanced and such gates can outperform LOQC equivalents. The technique
can be applied to other types of nonlinearities, making the implementation of
nonlinear optical gates much more attractive.
\end{abstract}


\maketitle
%==============================================================

%\structure{Introduction--- how optical systems are promising, but the interactions are hard.}
Encoding information on optical qubits is a promising route to quantum
information processing. In particular, by encoding the information in
polarisation at optical frequencies, a photon experiences essentially no
coupling to the environment in free space, but still can be manipulated
easily with passive linear optical devices. 

This strength of optical systems is also their weakness. Very weak coupling
to the environment means that it is also very difficult to get photonic
qubits to interact. Linear optical quantum computation (LOQC) \cite{0512071} is the most
developed scheme for optical quantum information processing, and it
sidesteps the interaction issue by using non-deterministic gates,
conditioned on photo-detection results. The inherent nondeterminism in LOQC is
hidden by making use of teleportation and encoding. Though
scalable in principle, such an approach takes a heavy penalty in
resources.

%\structure{Introduction--- to nonlinear schemes in general and to Fransons scheme in particular}
An alternative approach was recently proposed by Franson and co-workers
\cite{04franson062302, 06franson053817} for constructing an entangling gate. In this
proposal, two photonic qubits interact via evanescent coupling of the modes
between two fibre cores. Unwanted two photon terms are suppressed via the
Quantum Zeno effect \cite{77misra756} provided by strong two-photon absorption
within the fibre cores. A key question for the implementation of this gate is
the effect of single photon loss on it's function. Recent estimates
\cite{06franson053817} suggest that with appropriate engineering, the rate of two photon
absorption may be set four or five orders of magnitude larger than single
photon loss. Given such a large ratio of absorption rates, how does the gate
perform? Since manipulating the polarisation state of a single photon can be
currently achieved with high fidelity, the addition of an efficient entangling
gate would be a tremendous boost to optical quantum computation.

\begin{figure}[htb]
\centering
\includegraphics[width=.35\textwidth]{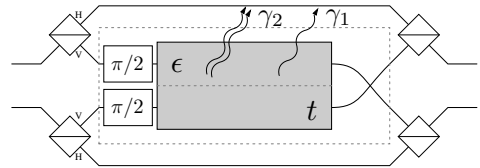}
\caption{Key parameters of the system. The
gate between the qubits is created by an interaction between the vertical
polarisation modes inside the device. The modes interact with a strength
$\epsilon$, for a time $t$. While passing through the medium the modes undergo
one- and two- photon loss with the rates $\gamma_{1}$ and $\gamma_{2}$,
respectively. The whole gate forms a dual-rail \textsc{csign} gate for
polarisation encoded qubits while the region inside the dashed box forms a
single-rail encoded \textsc{csign} gate.}
\label{fig:sketch}
\end{figure}

%\structure{what we contribute in this paper}
In this paper we evaluate a \textsc{csign} gate constructed using the Quantum
Zeno technique in the presence of photon loss --- a dominant decoherence
process in optical systems.  We model the gate by solving the master
equation evolution for the system, characterising the effect of the photon loss by calculating the
closeness of the gate to an ideal \textsc{csign} gate.
Finally, we propose
using some of the encoding protocols that have been developed for LOQC to
mitigate the effect of this loss. 

%\structure{describe our model}
A stylised representation of the gate is given in Fig.~\ref{fig:sketch}.  The
inner part of the gate (within the dashed rectangle in the figure) acts as a
\textsc{csign} on `single-rail' qubits, that is, on qubits where a logical
$\ket{0}_{L}$ is encoded as a vacuum state and a logical $\ket{1}_{L}$ as a
single Fock state. The two qubits interact with a strength $\epsilon$ within
the nonlinear region depicted, for a time $t$, and also undergo one- and
two-photon loss at rates $\gamma_1$ and $\gamma_2$, respectively.  Note that at
most there will be two photons within the gate corresponding to the $\ket{11}$
term. 

Since the ideal \textsc{csign} gate only changes the phase of the qubits and
not the photon number, the same gate can be
used for `dual-rail' encoded qubits. In dual-rail encoding there is one photon encoded in two
modes, such as with polarisation encoding where 
$\ket{0}_{L}\equiv\ket{H}=\ket{10}_{HV}$ and
$\ket{1}_{L}\equiv\ket{V}=\ket{01}_{HV}$, and the latter kets are Fock-state
occupation numbers.  The gate can be applied to the polarisation encoded qubits
also as depicted in Fig.~\ref{fig:sketch}.  The polarisation modes are spatially
split by a polarising beamsplitter and only the $V$ mode from each qubit goes
through the device which acts as a single-rail device for those modes, only
the $\ket{VV}$ term should acquire a phase-shift.  The polarisation modes are
recombined after the device.

%\structure{Write down the corresponding master equation. change variables to
%more convenient ones. Explain the new variables.}
The interaction between the two modes is described by the interaction
Hamiltonian for a beamsplitter $H_{I} = \epsilon (a_1^{\dagger}a_2+a_1 a_2^{\dagger})$ 
where $a_j^{\dagger}$, $a_j$ are the bosonic creation and annihilation operators respectively
for mode $j$. The interaction Hamiltonian transforms the modes as
$a_{1} \rightarrow \cos(\epsilon t) a_{1} +i \sin(\epsilon t)a_{2}$ and 
$a_{2} \rightarrow \cos(\epsilon t) a_{2} +i \sin(\epsilon t)a_{1}$.

In order to incorporate two photon absorption and single photon loss we model
the device by a damped master equation.  Following standard techniques, the
master equation for the system is
\begin{align}
\frac{d\rho}{d\tau} =& i [\rho,H_{I}] 
+ \frac{1}{2}\sum_j(2 a_j\rho a_j^{\dagger}-a_j^{\dagger}a_j\rho-\rho a_j^{\dagger}a_j)\nonumber \\ 
&+ \frac{\gamma}{2}\sum_j(2 a_j^2\rho a_j^{\dagger 2}-a_j^{\dagger 2}a_j^2\rho-\rho a_j^{\dagger 2}a_j^2),
\label{eq:master}
\end{align}
%\begin{align}
%\frac{d\rho}{dt} =& i [\rho,H_{I}] 
%+ \frac{\gamma_{1}}{2}\sum_j(2 a_j\rho a_j^{\dagger}-a_j^{\dagger}a_j\rho-\rho a_j^{\dagger}a_j)\nonumber \\ 
%&+ \frac{\gamma_{2}}{2}\sum_j(2 a_j^2\rho a_j^{\dagger 2}-a_j^{\dagger 2}a_j^2\rho-\rho a_j^{\dagger 2}a_j^2)
%\end{align}
where for convenience, we rescaled the interaction time as $\tau=\gamma_{1}t$ and
introduce the scaled interaction strength $\kappa=\epsilon/\gamma_{1}$ and
ratio of two-photon to single-photon loss $\gamma=\gamma_{2}/\gamma_{1}$.
In order to perform a  \textsc{csign} gate, we require $\kappa\tau=\pi/2$ which fixes the
value of $\kappa$, and now our system is governed by two parameters: $\tau$
which determines the total time spent in a medium which is characterised by the
ratio of loses $\gamma$.

%\structure{Reformulate the problem: we only have a restricted number of photons
%so expand in number states. Vectorise rho, which leads to coupled ODEs, now the
%problem looks familiar.} 
The key to solving Eq.~(\ref{eq:master}) is to observe
that there will be a maximum of only two photons in the system, corresponding
to the state $\ket{11}_{L}\equiv\ket{VV}$. We can do a complete expansion of
$\rho$ in the number-state basis
$\mathcal{B}\in\{\ket{00},\ket{01},\ket{10},\ket{11},\ket{02},\ket{20}\}$ as
there will be no other contributions, so $\rho = \sum_{ij,kl\in\mathcal{B}} d_{ijkl}\ketbra{ij}{kl}$.
We can write a differential equation for each component of the density matrix
by using $\bra{mn}\dot\rho\ket{pq}=\dot{d}_{mnpq}$ and
$\bra{mn}\rho\ket{pq}={d}_{mnpq}$. Applying these to Eq.~(\ref{eq:master}) we
arrive at the following coupled ODEs:
\begin{widetext}
\begin{equation}\label{eq:odes}
\begin{split}
\dot{d}_{mnpq} =& -i\kappa\left( \sqrt{m(n+1)}d_{(m-1)(n+1)pq}+
\sqrt{(m+1)n}d_{(m+1)(n-1)pq}-\sqrt{(p+1)q}d_{mn(p+1)(q-1)}
-\sqrt{p(q+1)}d_{mn(p-1)(q+1)}\right) \\
& + \sqrt{(m+1)(p+1)}d_{(m+1)n(p+1)q}+\sqrt{(n+1)(q+1)}d_{m(n+1)p(q+1)}\\
& + \gamma\left(\sqrt{(m+1)(m+2)(p+1)(p+2)}d_{(m+2)n(p+2)q}+\sqrt{(n+1)(n+2)(q+1)(q+2)}d_{m(n+2)p(q+2)} \right)\\
& - \frac{1}{2}\big( m+p+n+q+\gamma[m(m-1)+p(p-1)+n(n-1)+q(q-1)] \big)d_{mnpq},
\end{split}
\end{equation}
\end{widetext}
where we have used $a\ket{n}=\sqrt{n}\ket{n-1}$ and $a^{\dagger}\ket{n}=\sqrt{n+1}\ket{n+1}$.
The solutions $d_{mnpq}(\tau)$ to these equations are the matrix elements of $\rho$ written
in the $\mathcal{B}$ basis.

%\structure{Derive the solution to the ODEs: diagonalise, write solution,
%reconstruct rho. Discuss issues with existence of solutions?}
This system of first order linear ODEs can be written as a matrix equation,
if we take all the columns of $\rho$ and stack them on top of each other to make 
a vector $\vec{\rho}$, then $\frac{d}{d\tau} \vec{\rho}=A \vec{\rho}$.
Diagonalising the matrix $A=SDS^{-1}$ we can cast the problem in new
variables $\vec{\sigma}=S^{-1}\vec{\rho}$ which yield the solution
$\vec{\sigma}(t)=\exp(Dt)\vec{\sigma}(0)$. In terms of the original
density matrix we have $\vec{\rho}(t)=S\exp(Dt)S^{-1}\vec{\rho}(0)$.  We can then
re-shape the vector $\vec{\rho}$ to get $\rho(\tau)$

For the parameter choice $\kappa\tau=\pi/2$ the chosen interaction Hamiltonian
means that the modes transform as $a^{\dagger}\rightarrow -i b^{\dagger}$ and
$b^{\dagger}\rightarrow -i a^{\dagger}$ so that the ideal gate implemented is
almost a \textsc{swap} gate apart from some phases.  A \textsc{csign} gate can
be constructed from this gate $\mathbb{S}$ by introducing $\pi/2$ phases in each mode and
undoing the \textsc{swap} operation --- which can be done simply in optics by
directly swapping the two modes $\textsc{csign} = \textsc{swap} \;
\mathbb{S}\;\bigl(\begin{smallmatrix}1&0\\0&i\end{smallmatrix}\bigr) \otimes
\bigl(\begin{smallmatrix}1&0\\0&i\end{smallmatrix}\bigr)$.

\begin{figure}[htb]
\centering
\includegraphics[width=.47\textwidth]{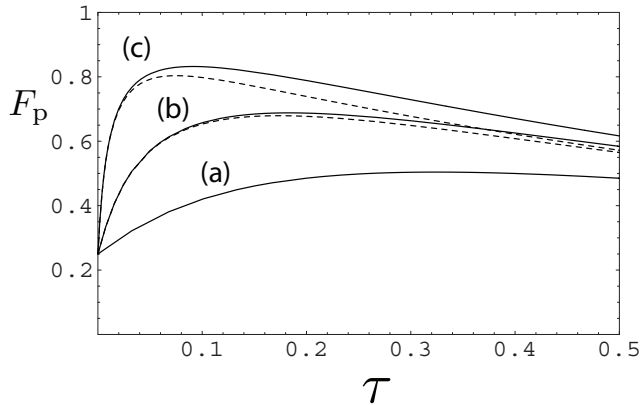}
\caption{Process fidelity of the gate with the ideal gate $\mathbb{S}$, for (a)
$\gamma=20$, (b) $\gamma=100$, and (c) $\gamma=500$. 
The dashed lines represent the 19 beam splitter case \cite{BSref}.
}
\label{fig:process-fidelity}
\end{figure}

We can quantify the gate performance by calculating the process fidelity
$F_{\mathrm{p}}$ of the gate with the ideal \textsc{csign} gate. A general quantum
operation $\mathcal{E}(\rho)$ on $\mathcal{H}$ is isomorphic to density
matrices in $\mathcal{H}\otimes\mathcal{H}$ \cite{72jamiolkowski275} via
$\rho_{\mathcal{E}} = \mathcal{I}\otimes\mathcal{E}(\ketbra{\phi}{\phi})$, 
where $\ketbra{\phi}{\phi}$ is a maximally entangled state $\sum_j \ket{j}\ket{j}/\sqrt{d}$ on the two
$\mathcal{H}$ spaces of dimension $d$.  The process fidelity is then the fidelity between these
process density matrices $F_{\mathrm{p}}=F(\rho_{\mathcal{E}},\rho_{\mathbb{S}})$.
The process fidelity is linearly related to the average gate fidelity $\bar{F}$,
via $F_{\mathrm{p}} = (\bar{F}d+1)/(d+1) $
where $d$ is the dimension of the system, so that $F_{\mathrm{p}}$ captures
the average performance of the gate \cite{99horodecki1888}, e.g. it has been shown that
$1-F_{\mathrm{p}}$ bounds the average probability of error in a function
computation \cite{05gilchrist062310}.
The process fidelity (which is identical whether we use single- or dual-rail
encoding) is plotted in Fig.~\ref{fig:process-fidelity} for a range of
$\gamma$. 
The most striking feature of the figure is that 
there is an optimum interaction time $\tau=\tau_\mathrm{opt}$ for a given
$\gamma$, and also that the maximum fidelity is not unity. The existence of this optimum is
intuitive --- too short an interaction time and there has been insufficient two
photon absorption for the gate to function, and too long an interaction time
and the effect of the single photon loss starts to dominate.
We will assume for the remainder of the paper that $\tau=\tau_{\mathrm{opt}}$
for the task at hand. 

This behaviour of the fidelity with photon loss is also seen in high efficiency
interaction free measurement \cite{99kwiat4725} which also makes use of the
quantum Zeno effect. In a nutshell, the high efficiency limit requires long
interaction times which enhances the effect of any loss.

%\comment{merge figures} 

\begin{figure}[htb]
\centering
\includegraphics[width=.47\textwidth]{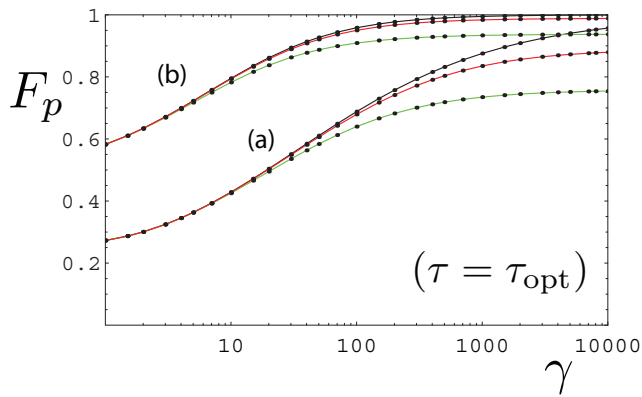}%{fidelitycz.eps}
\caption{(a) The scaling of the process fidelity with $\gamma$ for optimal
$\tau=\tau_{\mathrm{opt}}$ (separate optimisation for each point shown). 
(b) The process fidelity for the unbalanced encoded gate. Note
that attaining larger fidelities becomes increasingly difficult. The green and red lines represent the 7 and 19  beam splitter case, respectively \cite{BSref}.}
\label{fig:fidelity-scaling}
\end{figure}

In Fig.~\ref{fig:fidelity-scaling}~(a) we examine 
the scaling of the process fidelity with $\gamma$.
Note that though the final limit is unit
fidelity, it becomes increasingly difficult to attain higher process
fidelities, with the cost in $\gamma$ becoming super-exponential for high fidelities.

%\structure{Introduce parity+loss encoded qubits. How they work.}
The chief failure mode of the gate at long interaction times is photon loss.
If we could protect the qubits specifically against this type of loss we can
expect a significant boost to the fidelity of the gate. We can use a protocol
from LOQC for this task which has two levels \cite{05ralph100501}.  At the
first level the qubits are encoded in a parity code over $n$ physical
qubits. A parity $\ket{0}^{(n)}$ is a superposition of all the even parity
components, and a parity $\ket{1}^{(n)}$ is all the odd parity components:
\begin{align}
\ket{0}^{(n)} =& (\ket{+}^{\otimes n} + \ket{-}^{\otimes n})/\sqrt{2}\\
\ket{1}^{(n)} =& (\ket{+}^{\otimes n} - \ket{-}^{\otimes n})/\sqrt{2}
\end{align}
The key property of this code is that a measurement of one of the physical
qubits in the computational basis results in a heralded bit-flip on the encoded
qubit but does not measure the encoded qubit.  This code allows recovery from
teleportation failures in KLM style protocols, giving multiple attempt at
nondeterministic teleportation. Such codes have already been demonstrated
experimentally \cite{05obrien060303}.

If we know from which physical dual-rail qubit the photon has been lost, the
parity code translates photon loss into an unheralded bit-flip on that qubit.
Photon loss acts as if the environment had measured the qubit but we do not
have access to the result so the state is left in a mixture of obtaining both
measurement outcomes.
In order to recover from this unheralded bit-flip a simple repetition code suffices
and the full encoding becomes:
\begin{equation}
\alpha\ket{0}_L+\beta\ket{1}_L = \alpha\ket{0}^{(n)}\ket{0}^{(n)}\dots+\beta\ket{1}^{(n)}\ket{1}^{(n)}\dots 
\end{equation}
where $q$ parity qubits are tensored together. The code can be optimised by carefully 
choosing the values of $n$ and $q$. In this paper, we will assume that the only source of loss 
is the nonlinear interaction so that a value of $n=2$ will be sufficient, so that the 
basic parity qubits will be Bell states: $\ket{0}^{(2)}=(\ket{00}+\ket{11})/\sqrt{2}$ and 
$\ket{1}^{(2)}=(\ket{01}+\ket{10})/\sqrt{2}$.

%\structure{Gate performance on encoded qubits, doing a \textsc{csign} on
%encoded qubit.}
To perform a \textsc{csign} on the encoded qubits we attempt the gate on two
photons, one from each qubit, then immediately measure the qubits in the
$\ket{\pm}$ basis and determine if a photon was lost or not.  If we
successfully obtain a result from the measurements (i.e. no photon was lost) we
then measure the remaining parity qubit in the computational basis, resulting in a
\textsc{csign} being implemented on the remainder of the code (local phase-flips may need to be
applied depending on the measurement results).  On the other hand, if one or
both detectors fail to register an event then we remove the affected part of
the redundancy code by measuring the remaining parity qubit in the $\ket{\pm}$
basis (again phase-flip corrections may need to be applied). 

Applying the gate to encoded qubits and considering the
conditional case where no photons have been lost (in the case of loss we
recover the qubits and try again) we obtain the fidelity in
Fig.~\ref{fig:fidelity-scaling}~(b). This fidelity also has an optimum value
with $\tau$ after which it reduces with larger $\tau$. The reason is that
loss from the gate is unbalanced: the $\ket{HH}$ component suffers no loss
while a component with $\ket{V}$ suffers loss. In the long $\tau$ limit
states get skewed towards $\ket{HH}$ reducing the average fidelity.  We can
`balance' the gate by also including single photon loss in the $\ket{H}$ arms
of Fig.~\ref{fig:sketch}, in this case the gate has an optimum fidelity of 1 
(in the long $\tau$ limit).

%\structure{Resource requirements--- we're offsetting loss tolerance against
%resource usage. We should quantify this here.}
Increasing the fidelity comes at a price --- each attempt of the gate consumes
four qubits from the code. In the cases where we detect photon loss the
protocol resets the qubits to their original state and we need to attempt the
gate again.  The lower the probability of success the more qubits we consume on
average per successful gate operation. We need to balance the trade-off between
fidelity of the gate operation and the amount of resources consumed.

In the unbalanced case, the success probability is also state-dependent. 
%\comment{is it state independent in the balanced case?} 
Since the polarisation encoded gate
acts on only the $\ket{V}$ component and we are modelling loss only in the
nonlinear device, the maximum loss is suffered by the $\ket{VV}$ component,
while the $\ket{HH}$ component suffers no loss. This means that detecting a
photon loss event will do a partial measurement of the qubits in the
computational basis, but this does not pose a problem since the parity code is
specifically designed for this scenario.

We can quantify the success probability in both the balanced and unbalanced cases by the average 
probability of success
\begin{equation}
\bar{p} = \sum_{ab} \int\!\! d\psi\; \mathrm{Tr}\{ \ketbra{ab}{ab} \mathcal{E}(\psi)  \},
\end{equation}
where the integral is over the Haar measure for all two-qubit pure states $\psi$,
and $a,b \in \{+,-\}$ are the successful measurement results after the gate.
Because of the linearity of the trace and the operation this reduces to
$\bar{p} = \mathrm{Tr}\{ I_{(2)} \mathcal{E}(I_{(2)}/4)\}$,
where $I_{(2)}$ is the identity in the two-qubit subspace where the photons have not been lost.

%\structure{comparison with LOQC}
Say we are given two logical qubits encoded over some number of photons and the
goal is to produce a \textsc{cnot} or \textsc{csign} gate with a process
fidelity $F_{\mathrm{p}} \ge F_{\mathrm{p,min}}$, and a probability of catastrophic
failure $P_{\mathrm{f}}\le P_{\mathrm{f,max}}$ (we lose all the photons and hence the logical
qubits).  We can compare against a competing LOQC protocol such as the gate
described in \cite{0505125} which acts on parity encoded qubits.  Given optimal
encoding in both protocols, at what value of $\gamma$ does the quantum Zeno
gate begin to use fewer resources than the  LOQC gate?

Assuming that the
target fidelity can be met automatically, the LOQC gate's performance is
governed entirely by $P_{\mathrm{f,max}}$ and this determines the minimum size of the
parity code needed.  The probability of operation per attempt is $1/4$ and so
$P_f=(3/4)^{n/2}$. Since each attempt consumes 3 photons (one each from
control, target and resource) and all the control qubits photons have to be
measured out at the end, the average number of photons consumed per successful
gate operation will be $4\!\times\! 2\!+\!n/2$.
For the gate presented in this paper, achieving the target fidelity is determined by the ratio of
absorptions $\gamma$, which in turn determines the probability of success
$P_{\mathrm{s}}=1-P_{\mathrm{f}}$, and hence the average number of photons consumed per successful gate
operation is $4/P_{\mathrm{s}}$ (we consume four photons per attempt). The probability
of catastrophic failure is determined by the size of the code and is
independent of the consumption since we do not have to re-encode for a
successful gate operation as in \cite{0505125}.

If the target fidelity is $F_{\mathrm{p,min}}=99.9\%$ then a $\gamma$ of around 4000 is needed for
the unbalanced gate, but this gate consumes 7 times \emph{less} resources than the
LOQC scheme for $P_{\mathrm{f,max}}=0.01$ and 15 times less resources for
$P_{\mathrm{f,max}}=0.0001$. By using the balanced version, $\gamma$ can be traded off
against probability (and hence resource consumption) so for example, at
$\gamma=500$ and $P_{\mathrm{f,max}}=0.01$ the Zeno gate still consumes around five times less
photons.

%\structure{test-bed experimental proposal}

%\structure{Conclusion}
We have investigated the performance of the quantum Zeno gate in the presence
of single photon absorption. Under this noise the raw gate fidelity is severely
hampered, even when the nonlinear two photon absorption rate is some four
orders of magnitude larger than the single photon loss rate.  However, if such
a gate was used within a context of error correction, we have found that higher
gate performance can be achieved at the cost of consuming a resource in the
form of the logical qubit encoding. Even more promising, there are parameter
regions where a high fidelity is achieved with high efficiency, and the
resource consumption is less than for corresponding LOQC protocols. 
Also by restricting the gate operation to certain tasks \cite{0609224} it is likely that
further savings could be achieved.
It should be noted that the technique presented here is not specific to the quantum Zeno gate 
and could be used to mitigate loss in other nonlinear gates.
Being able to tolerate some amount of photon loss in its operation makes the
practical design of such gates far less stringent and consequently more
attractive.

%\structure{Acknowledgements}
%% Acknowledgements
We would like to thank Andrew White, Andrew Doherty and Tim Ralph for helpful
discussions. CRM is supported in part by NSERC, ARO, CIAR, MITACS and the
Lazaridis Fellowship, and AG is supported by the Australian Research Council and the DTO-funded U.S.
Army Research Office Contract No. W911NF-05-0397.

%\bibitem[$*$] {$*$} alexei@physics.uq.edu.au
%\bibliography{references.bib}

\begin{thebibliography}{13}
\bibitem[{BSr()}]{BSref}
\bibinfo{note}{As an alternative we can model our \textsc{csign} as an array of
  partially transmitting beam splitters with one- and two-photon loss in
  between each. We can solve for the output of the $n$'th beam splitter/loss
  combination by solving a set of recurrence equations along with Eq.
  (\ref{eq:odes}) with $\kappa=0$.}
\end{thebibliography}

\end{document}